\documentclass[journal=jacsat,manuscript=article]{achemso}

\usepackage{amsmath}
\usepackage{graphicx}
\usepackage[colorlinks=true, allcolors=blue]{hyperref}
\usepackage{physics}
\usepackage{float}
\usepackage{enumerate}
\usepackage{amsfonts}
\usepackage{caption}
\usepackage{subcaption}
\usepackage{bm}
\usepackage{cleveref}
\usepackage{booktabs} 
\usepackage{array}
\usepackage{longtable}
\usepackage{newtxtext,newtxmath}
\usepackage{dcolumn}
\usepackage{epstopdf}
\usepackage{algorithm}
\usepackage{algpseudocode}

\title{Solve Crude Oil Scheduling Problems by Using Quantum-Classical Hybrid Algorithms}

\author{Jian Yang}
\email{yangjian2020@petrochina.com.cn}
\author{Bohang Wang}
\author{Lina Wang}
\author{Jiacheng Chen}
\affiliation[PetroChina]
{Petrochina (BEIJING) Digital Intelligence Research Institute CO., LTD, 102206, Beijing, China}

\author{Gaoxiang Tang}
\affiliation[Tsinghua University]
{Center for Quantum Information, IIIS, Tsinghua University, 100084, Beijing, China}

\author{Zihan Deng}
\affiliation[Huayi Boao Beijing]
{Huayi Boao (Beijing) Quantum Technology Co., Ltd., 100176, Beijing, China}

\author{Wending Zhao}
\email{zhaowending0525@gmail.com}
\affiliation[Huayi Boao Beijing]
{Huayi Boao (Beijing) Quantum Technology Co., Ltd., 100176, Beijing, China}
\alsoaffiliation[Huayi Boao Hunan]
{Huayi Boao (Hunan) Quantum Technology Co., Ltd., 410000, Hunan, China}
\alsoaffiliation[BIT Interdisciplinary]
{School of Interdisciplinary Science, Beijing Institute of Technology, 100081, Beijing, China}
\alsoaffiliation[BIT Lab]
{Beijing Key Laboratory of Quantum Matter State Control and Ultra-Precision Measurement Technology, 100081, Beijing, China}

\author{Xianfeng Cai}
\email{chaixianfeng@petrochina.com.cn}
\affiliation[PetroChina]
{Petrochina (BEIJING) Digital Intelligence Research Institute CO., LTD, 102206, Beijing, China}

\begin{document}

\begin{abstract}
The optimization of front-end crude oil scheduling is a critical determinant of refinery profitability and operational stability. However, the coupling of discrete logistics events (e.g., vessel berthing) with continuous material flows (e.g., pipeline transfers) renders this problem an NP-hard Mixed-Integer Linear Programming (MILP) challenge, often intractable for classical solvers at industrial scales. This study proposes a novel hybrid quantum-classical framework to address these computational bottlenecks. We employ Benders Decomposition to decouple the monolithic model into a discrete Master Problem (MP) and a continuous Subproblem (SP). To exploit the search capabilities of quantum computing, the MP is reformulated as a Quadratic Unconstrained Binary Optimization (QUBO) model and solved via a hybrid quantum solver, while the SP enforces mass balance and quality constraints through iterative optimality and feasibility cuts. Extensive experiments on 15 multi-scale instances demonstrate that the proposed framework significantly outperforms traditional metaheuristics (e.g., Genetic Algorithms, Tabu Search), reducing total operating costs by approximately 73--80\% and achieving computational speeds comparable to state-of-the-art commercial solvers (Gurobi). By effectively leveraging global optimality cuts, the method overcomes the tendency of heuristic approaches to trap in local optima, providing a robust and scalable solution for complex refinery logistics.
\end{abstract}

\section{Introduction}

In the global energy architecture, the refining industry serves as the critical link between crude oil resources and final refined products, including gasoline, diesel, and petrochemical feedstocks. Given the sustained growth in global demand for petroleum and petrochemical products \cite{citaristi2022international,MarketReport2017}, crude oil scheduling optimization has emerged as a pivotal element in the energy and chemical industry chain, whose efficiency profoundly impacts energy supply security, economic performance, and environmental sustainability. This optimization problem encompasses not only the coordinated planning of upstream crude oil procurement, transportation, and extraction but also directly influences the efficiency and quality of downstream refining operations and petrochemical production \cite{kelly2003crude,kelly2003crudePart2,jia2004efficient}.

In response to declining profit margins due to intense market competition and the increasing growth of operating costs, refineries have an increasingly significant demand for enterprise-level optimization systems\cite{khor2017petroleum}. Optimizing crude oil scheduling systems helps reduce costs, minimize inventory and environmental impact, while maximizing profitability and responsiveness\cite{grossmann2012advances,pinto2000planning}. An efficient scheduling system can enhances crude oil transmission efficiency, increases the sulfur content of the processed crude, and improves overall product quality, thereby reducing operating costs or increasing profitability. For example, Kelly et al. \cite{kelly2003crude,kelly2003crudePart2} optimized such a system in a refinery with a capacity of 100,000 barrels per day, raising the sulfur content from $0.55\%$ to $0.85\%$, which resulted in an additional 2 million dollars in annual profit while simultaneously reducing operating expenditures.

The crude oil scheduling problem is intrinsically a batch process scheduling problem. It is generally depicted using topological network structures, primarily divided into the State-Task-Network (STN) method \cite{kondili1993general} and the Resource-Task-Network (RTN) method \cite{pantelides1994unified}. Mathematically, these are abstracted into combinatorial optimization problems(COPs), often expressed as Mixed-Integer Linear Programming (MILP) or Mixed-Integer Nonlinear Programming (MINLP) models, depending on the treatment of discrete events and continuous flow constraints \cite{lee1996mixed, jia2003refinery}. Unfortunately, precisely solving such problems poses significant challenges. The accuracy and computational time required for a solution are proportional to the solution space, which increases exponentially with the degree of discretization and the number of variables(NP-hard). This makes strict mathematical solutions computationally infeasible for meeting the high concurrency and rapid feedback demands of real-world business operations\cite{shenoy1980dynamic, kochenberger2014unconstrained, riandari2021quantum}.

To overcome the aforementioned challenges, classical computing often employs heuristic algorithms and data-driven machine learning algorithms to achieve rapid approximate solutions. Heuristic algorithms, such as Simulated Annealing \cite{alkhamis1998simulated}, Tabu Search \cite{glover1998adaptive}, and Genetic Algorithms \cite{goldberg1989messy}, have been widely applied to obtain feasible solutions within reasonable timeframes \cite{motlaghi2008expert,ramteke2012large,venter2016solution}. However, these methods are prone to being trapped in local optima, making it challenging to ensure solution quality.
In recent years, data-driven machine learning algorithms have been widely applied to solving combinatorial mathematical optimization problems, such as using supervised learning for branching strategies in Branch-and-Bound algorithms \cite{marcos2014supervised,zhang2020gbdt,he2014learning37} and employing Reinforcement Learning (RL) and Graph Neural Networks (GNN) for direct scheduling decisions \cite{gasse2019exact,chen2023reinforcement,qin2025crude}. While ML approaches reduce online solving time, they often require extensive training data and may still struggle with the combinatorial explosion in ultra-large-scale instances. 

Quantum Computing (QC), utilizing quantum superposition and entanglement to provide theoretical acceleration and optimization, has emerged as a disruptive technology for solving COPs\cite{albash2018adiabatic, preskill2018quantum}. 
Currently, quantum algorithms such as Quantum Annealing \cite{kadowaki1998quantum}and the Quantum Approximate Optimization Algorithm (QAOA)\cite{farhi2014quantum} are applied to solving Quadratic Unconstrained Binary Optimization (QUBO) problems, and demonstrate the potential advantage in solving NP-hard scheduling problems\cite{venturelli2015job, yarkoni2022quantum}. However, despite the rapid advancement of quantum combinatorial optimization algorithms in process systems engineering \cite{ajagekar2019quantum, bernal2022perspectives}, the application of quantum computing to complex industrial scheduling problems remains in the exploratory phase, especially crude oil scheduling problems. The unique constraints of refinery operations, such as component concentration consistency and mass balance in complex pipeline networks,pose specific challenges for quantum algorithm, which is modeled by binary variable encoding and constraint softening with slack variables \cite{lucas2014ising, glover2018tutorial}.

To address these challenges, this paper proposes a novel quantum-classical hybrid optimization framework tailored for the crude oil scheduling problem. Utilizing Benders decomposition, the complex Mixed-Integer Linear Programming (MILP) model is decoupled into a binary Master Problem (MP), which manages discrete scheduling events, and a linear Subproblem (SP), which handles continuous material flows. To leverage the ability of quantum computing, the MP is reformulated into a QUBO problem and solved via a hybrid solver\cite{zhao2025clustering}, while the SP generates optimality and feasibility cuts to guarantee global physical constraints such as mass balance and component concentration. This architecture effectively bridges the gap between quantum capabilities and complex industrial constraints. Extensive experiments on 15 scale-varying instances demonstrate that the proposed method significantly outperforms traditional metaheuristics (e.g., Genetic Algorithm, Tabu Search) in both solution quality and convergence speed, successfully avoiding local optima where classical methods falter.

\section{Description and Modeling of Crude Oil Scheduling Problem}
\label{sec:formulation}

The crude oil scheduling is the critical interface between maritime crude oil supply and the continuous production of refineries. The primary task involves coordinating material distribution, inventory control, and pipeline scheduling from oil tankers to Crude Distillation Units (CDUs) over a discrete time horizon \cite{lee1996mixed}. Tanks can be dedicated or used for storing crude oil mixtures. The final crude oil mixtures are blended in feed tanks to meet quality specifications before being sent to the CDUs for processing. Due to the coupling of discrete logistics events (e.g., vessel berthing) and continuous material flows (e.g., pipeline transfer rates), this problem is characterized as a large-scale combinatorial optimization problem constrained by limited resources, strict time windows, and quality specifications \cite{jia2004efficient, li2007improving}.

\begin{figure}[htbp]
    \centering
    \includegraphics[width=0.9\columnwidth]{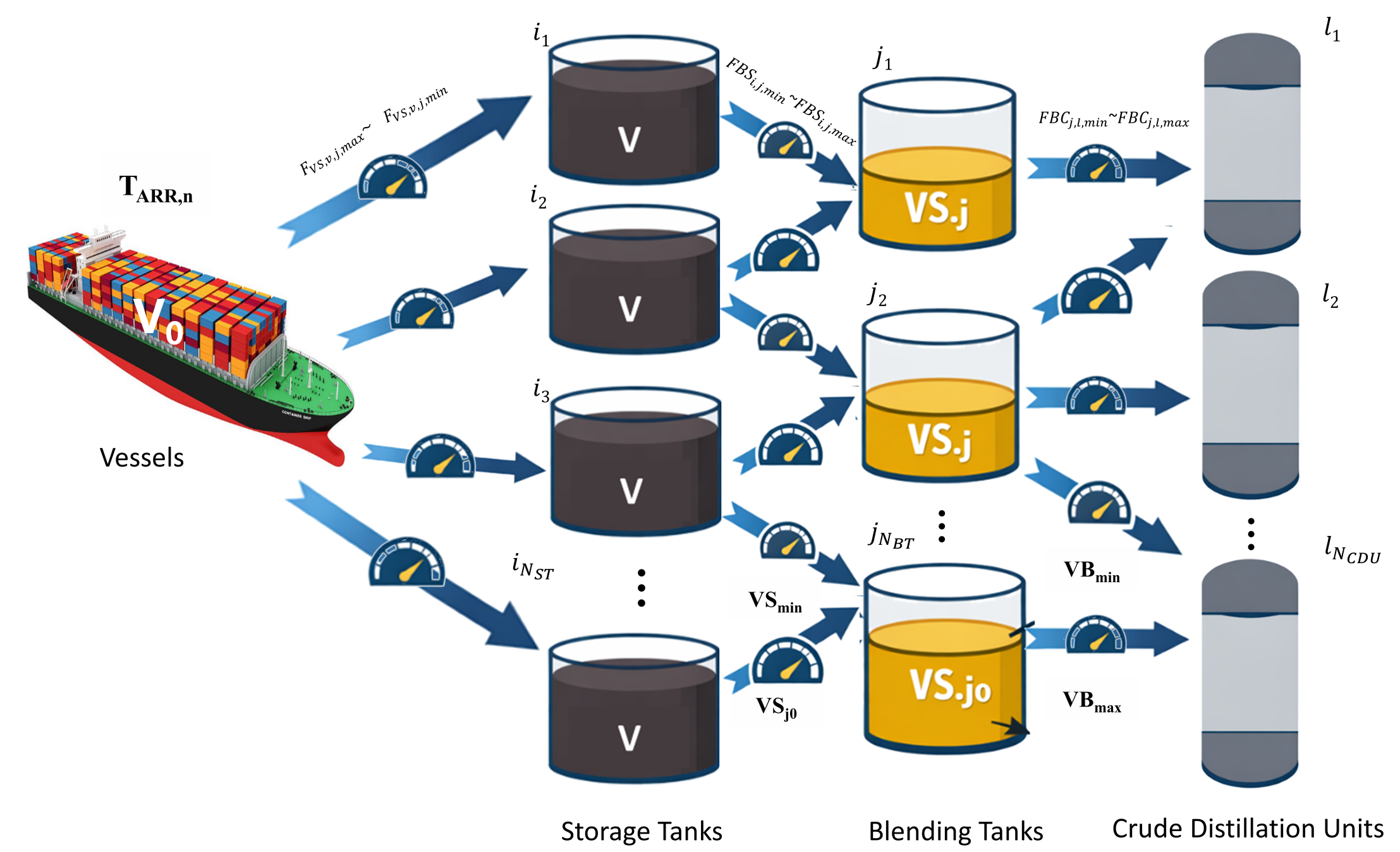}
    \caption{Illustration of the crude oil blend scheduling system and network.}
    \label{fig:1}
\end{figure}

\subsection{Crude Oil blend Scheduling System Introduction}

The physical topology of the system is modeled as a multi-stage network comprising four primary entity levels, as illustrated in Figure \ref{fig:1}. In summary, the process of unloading oil from tankers and subsequent distillation production can be classified into four stages:

\begin{itemize}
    \item \textbf{Vessels ($V$)}: These are the material sources of the system. Each vessel carries specific types of crude oil and is scheduled to arrive at the port within a predetermined time window. Due to the limited availability of berthing resources, such as jetties and unloading pipelines, vessels frequently encounter congestion.  If a vessel's retention time—defined as the duration from arrival to departure—exceeds the contractually allowable laytime, the refinery incurs significant demurrage costs.
    
    \item \textbf{Storage Tanks ($i$)}: Directly connected to the unloading pipelines at the dock, these primary tanks act as the first buffer pool. They receive crude oil discharged from vessels. The management of these tanks is constrained by physical capacity limits and the need to segregate crude types to prevent undesirable mixing at this initial stage. Simultaneously, storage tanks also function as crude oil storage units, facilitating the unloading of crude oil from vessels, thereby reducing costs.
    
    \item \textbf{Blending Tanks ($j$)}: Located downstream of the storage tanks, these secondary tanks function as the "feeding" buffer.  They receive crude oil from storage tanks for necessary blending, settling, or water drainage before the material is fed to the production units. These tanks are critical control nodes where crude properties are homogenized to meet the specific diet requirements of the CDUs.
    
    \item \textbf{Crude Distillation Units ($l$)}: The material sinks of the system are the crude distillation units (CDUs), which operate continuously and necessitate a stable feed rate and consistent feedstock quality. Any fluctuation in feed properties or a disruption in supply can cause significant operational instability, pose safety risks, and result in economic losses. Consequently, the scheduling model must regard CDU demand ($DM$) as a strict constraint that must be met. These constraints also introduce challenges in solving the optimization problem.
    
\end{itemize}

\subsection{The Mathematical Formulation of Cost function and Constraints}

The mathematical formulation relies on two distinct layers of coupled decision variables that jointly determine schedule feasibility and economic performance. Discrete binary scheduling Variables (binary) determine the timing and status of operations, such as assigning vessel berthing windows and managing dynamic pipeline connectivity (e.g., switching connections between charging tanks and CDUs), where minimizing switching frequency is crucial for system stability. Simultaneously, Continuous Logistics Variables determine the magnitude of material flows and inventory levels based on these active connections, ensuring mass balance across the network so that total unloading volumes equal cargo limits and tank levels respect safety bounds defined by inflows and outflows.

In the optimization of refinery crude-oil scheduling, the key objective is to minimize total costs over the planning period while satisfying the constraints of pipelines and tanks. In this work, the cost function composed of the following four parts:
\begin{enumerate}
    \item \textbf{Demurrage Costs}: A penalty function that increases linearly with the delay in vessel operations (see Eq.~\ref{eq:obj_dem_fix}). 
    This drives the schedule to prioritize the timely berthing and unloading of arriving vessels. 
    However, excessively early and rapid scheduling of crude oil unloading can disrupt the crude oil composition ratio in the mixing tank, thereby reducing production efficiency. This disruption introduces significant challenges to the optimization of the scheduling process.
    
    \item \textit{Fixed Unloading Costs}: Operational fees associated with the setup and usage of port facilities for discharging cargo (see Eq.~\ref{eq:obj_dem_fix}).
    This type of cost is typically proportional to the volume of crude oil transported and exerts minimal influence on the optimization strategy for scheduling.
    
    \item \textit{Pipeline Setup Costs}: Penalties applied to switching operations in the pipeline network (see Eq.~\ref{eq:obj_setup}). 
    Given the constraints inherent in storage and blending tanks, switching operations are often unavoidable and thus significantly influence the optimization of the final scheduling plan. Reducing the frequency of tank switch-overs will minimize these costs and promote stability in the feeding configuration.
    
    \item \textit{Inventory Holding Costs}: Costs derived from the cumulative volume of crude oil held in storage and charging tanks (see Eq.~\ref{eq:obj_hold}). 
    This term reflects the capital tied up in inventory and incentivizes the efficient throughput of material, although it must be balanced against the need for safety stock.
    
\end{enumerate}

Actual refinery scheduling incurs additional costs beyond these cost function above. Nevertheless, this omission does not compromise the general applicability of our quantum-classical hybrid algorithm. The detailed mathematical expression of the objective function is provided in App.~\ref{app:obj_func}.

To ensure schedule feasibility and economic performance, the decision variables and objective function are constrained by a strict set of conditions (see App.~\ref{app:constraints}). These constraints reflect the refinery’s physical limitations and supply-demand balance and can be broadly classified into three categories:
\begin{enumerate}
    \item \textit{Scheduling Logic and Time Windows} (Eq.~\ref{eq:con_unique} -- \ref{eq:con_tw}): These constraints govern the discrete events of the system, ensuring that vessels operate within their allowable arrival and departure windows, berth resource limits are not exceeded, and continuous unloading duration requirements are logically met.
    \item \textit{Cumulative Inventory and Material Balance} (Eq.~\ref{eq:con_vessel_bal} -- \ref{eq:con_cdu}): Operating without explicit inventory state variables, the model enforces mass conservation by bounding the cumulative material flows. These constraints ensure that total unloaded volumes equal cargo limits, tank capacities remain strictly within safety bounds at all time steps, and CDU production demands are fully satisfied.
    \item \textit{Operational Constraints} (Eq.~\ref{eq:con_unload_conn} -- \ref{eq:con_transfer_conn}): These bridge the discrete scheduling logic and continuous material flows, dictating that material transfer can only occur when physical pipeline connectivity is active and ensuring that flow rates respect the minimum and maximum pumping capacity limitations.
\end{enumerate}

\subsection{Mathematical Modeling for MILP Formulation}

To solve this optimization problem using standard mathematical programming solvers, we consolidate the decision variables and constraints defined previously into a compact Mixed-Integer Linear Programming (MILP) format.

First, we aggregate the distinct decision variables into two vectors:
\begin{itemize}
    \item Let $x \in \{0,1\}^n$ represent the vector of all \textbf{discrete variables}, collecting the vessel states and pipeline connection states. The dimension $n$ corresponds to the total number of discrete decision points over the horizon $T$.
    \item Let $y \in \mathbb{R}^p_{+}$ represent the vector of all \textbf{continuous variables}, collecting all material movement rates. The dimension $p$ corresponds to the total number of flow arcs in the network over time $T$.  
\end{itemize}

Accordingly, the objective function parameters are mapped to coefficient vectors $c$ and $h$. The vector $c$ captures all event-triggered costs (demurrage, fixed unloading, and setup fees), while $h$ encapsulates the inventory holding cost coefficients derived from the time-weighted flow formulation in Eq. (4).

The constraints are classified based on the variables they involve:
\begin{itemize}
    \item \textbf{Coupling Constraints} ($Ax + Gy \le b$): These represent the hybrid restrictions where continuous flows are bounded by discrete states (e.g., Big-M constraints Eq. 12-14) or linked via mass conservation (Eq. 10-11). Here, matrices $A$ and $G$ encode the network topology and capacity limits.
    \item \textbf{Logic Constraints} ($Bx \le b'$): These strictly govern the discrete scheduling rules (Eq. 5-9), such as time windows, sequence enforcement, and resource exclusivity, involving only the binary vector $x$.
\end{itemize}

The resulting general MILP formulation is expressed as follows:
\begin{align}
\max_{x,y}\quad & c^\top x + h^\top y \\
\text{s.t.}\quad & A x + G y \le b, \label{eq:AG}\\
& B x \le b', \label{eq:Bx}\\
& x \in \{0,1\}^n,\quad y \in \mathbb{R}^p_{+}.
\end{align}

\section{Hybrid Quantum–Classical Approach for Solving the MILP}
\label{sec:method}

\subsection{Benders Decomposition: From the MILP to a Continuous LP Subproblem and a Binary Master}
\label{subsec:benders}

The MILP in Sec.~\ref{sec:formulation} mixes binary scheduling decisions with continuous flow decisions, which makes direct solution expensive. Benders decomposition~\cite{naghmouchi2024mixed} separates these two roles. The idea is to keep only the binary variables in a master problem (MP) and push all continuous variables into a linear programming subproblem (SP). The two are then linked by cuts that are learned iteratively. Intuitively, the MP proposes a binary schedule $x$; given this $x$, the SP checks whether feasible flows exist and how much value they can deliver. The SP (or its dual) returns either a performance certificate (an \emph{optimality cut}) that tightens the MP’s prediction for the flow value, or a feasibility certificate (a \emph{feasibility cut}) that rules out the current $x$ because no flows could realize it. This procedure guarantees global optimality for the original MILP because the standard Benders assumptions are met.

Formally, introduce the residual right-hand side $d(x):=b-Ax$ and define, for any fixed $x\in\{0,1\}^n$, the primal subproblem
\begin{equation}
\label{eq:sp}
\begin{aligned}
\mathrm{SP}(x):\quad \max_{y}\ & h^\top y \\
\text{s.t.}\ & G y \le d(x),\quad y\ge 0.
\end{aligned}
\end{equation}
This is a linear program. Its dual is
\begin{equation}
\label{eq:spd}
\begin{aligned}
\mathrm{SPD}(x):\quad \min_{\mu}\ & d(x)^\top \mu \\
\text{s.t.}\ & G^\top \mu \ge h,\quad \mu \ge 0.
\end{aligned}
\end{equation}
Because both \eqref{eq:sp} and \eqref{eq:spd} are LPs, strong duality applies: if \eqref{eq:sp} is feasible and bounded, then $\max \eqref{eq:sp}=\min \eqref{eq:spd}$ and an optimal dual solution $\mu^\star$ exists. 
The dual outcome determines which cut is added to the master problem. If \eqref{eq:spd} is feasible with value $d(x)^\top\mu^\star$, an \emph{optimality cut} is introduced:
\begin{equation}
\label{eq:optcut-add}
\phi \;\le\; \mu^{\star\top} b \;-\; (A^\top \mu^\star)^\top x,
\end{equation}
which tightens the master’s prediction of $\phi$ whenever $d(x)^\top\mu^\star < \phi$ (within a tolerance). On the other hand, if \eqref{eq:spd} is infeasible, a Farkas certificate $r\ge 0$ with $G^\top r\ge 0$ and $d(x)^\top r<0$ proves that no feasible $y$ exists for the current $x$. In that case, a \emph{feasibility cut} is added:
\begin{equation}
\label{eq:feascut-add}
(A^\top r)^\top x \;\le\; r^\top b,
\end{equation}
which eliminates the infeasible schedule and any others that would violate the same structural condition.

Let $\phi$ denote the optimal value of $\mathrm{SP}(x)$. Using $\phi$ as an epigraph variable gives a master problem that retains only the binaries:
\begin{equation}
\label{eq:mp}
\begin{aligned}
\max_{x,\phi}\ & c^\top x + \phi \\
\text{s.t.}\ & Bx \le b', \\
& \phi \le \mu^\top b - (A^\top \mu)^\top x,\quad \forall\,\mu\in\mathcal{O}, \\
& (A^\top r)^\top x \le r^\top b,\quad \forall\,r\in\mathcal{F}, \\
& x\in\{0,1\}^n,
\end{aligned}
\end{equation}
where $\mathcal{O}$ collects optimality cuts and $\mathcal{F}$ collects feasibility cuts. Each optimality cut arises from a dual solution $\mu$ of \eqref{eq:spd} and has the form
\begin{equation}
\label{eq:opt-cut}
\phi \ \le\ \mu^\top b - (A^\top \mu)^\top x,
\end{equation}
which upper-bounds the best achievable flow value $\phi$ for \emph{all} schedules $x$ (it is a supporting hyperplane to the SP value function). Each feasibility cut arises from a Farkas certificate $r\ge 0$ with $G^\top r\ge 0$ and $d(x)^\top r<0$, proving that \eqref{eq:sp} would be infeasible under some $x$; the cut
\begin{equation}
\label{eq:feas-cut}
(A^\top r)^\top x \ \le\ r^\top b
\end{equation}
then excludes that schedule and any others that would repeat the same infeasibility mechanism.

\begin{algorithm}[t]
\caption{Classical Benders loop for the MILP (as implemented in \texttt{benders\_decomposition})}
\label{alg:benders}
\begin{algorithmic}[1]
\Require $A,G,b,B,b',c,h$, tolerance $\varepsilon>0$, maximum iterations $K$
\State Initialize cut pools $\mathcal{O}\gets\emptyset$, $\mathcal{F}\gets\emptyset$
\State Compute a safe $\phi_{\max}$ by solving the linear relaxation of Sec.~\ref{sec:formulation} with $0\le x\le 1$, $y\ge 0$
\State Set $UB\gets -\infty$, $LB\gets -\infty$
\For{$k=1,2,\dots,K$}
  \State \textbf{Master solve:} solve \eqref{eq:mp} with $\phi\le \phi_{\max}$ to get $(x^{(k)},\phi^{(k)})$; set $UB\gets \max\{UB,\,c^\top x^{(k)}+\phi^{(k)}\}$
  \State Form $d^{(k)} \gets b - A x^{(k)}$
  \State \textbf{Dual SP:} solve \eqref{eq:spd}; if feasible, obtain $(\mu^{(k)},\,d^{(k)\top}\mu^{(k)})$
  \If{\eqref{eq:spd} feasible \textbf{and} $d^{(k)\top}\mu^{(k)} < \phi^{(k)} - \varepsilon$}
     \State add optimality cut $\phi \le \mu^{(k)\top} b - (A^\top \mu^{(k)})^\top x$ to $\mathcal{O}$
     \State $LB\gets \max\{LB,\,c^\top x^{(k)} + d^{(k)\top}\mu^{(k)}\}$; \textbf{continue}
  \ElsIf{\eqref{eq:spd} infeasible}
     \State \textbf{Farkas ray:} solve $\min_r d^{(k)\top} r$ s.t. $G^\top r\ge 0$, $r\ge 0$, $\mathbf{1}^\top r\le 1$
     \If{$d^{(k)\top} r < 0$} \State add feasibility cut $(A^\top r)^\top x \le r^\top b$ to $\mathcal{F}$; \textbf{continue} \Else \State \textbf{stop} with status ``stalled'' \EndIf
  \Else
     \State \textbf{converged} if $\lvert \phi^{(k)} - d^{(k)\top}\mu^{(k)} \rvert \le \varepsilon$ (strong duality)
     \State optionally solve the primal SP \eqref{eq:sp} to recover $y^{(k)}$
     \State \textbf{return} $(x^{(k)}, \phi^{(k)}, y^{(k)})$, bounds $(UB,LB)$
  \EndIf
  \State \textbf{(Optional tightening)} update $\phi_{\max}\gets \min_{(\alpha,\beta)\in\mathcal{O}}\{\alpha - \sum_j \max(\beta_j,0)\}$, where $\alpha=\mu^\top b$, $\beta=A^\top\mu$
\EndFor
\State \textbf{return} best-known incumbent and bounds if $K$ reached
\end{algorithmic}
\end{algorithm}

\subsection{QUBO Reformulation of the Master Problem}
\label{subsec:mp_to_qubo}
We map the MP into an unconstrained quadratic binary optimization (QUBO) to enable quantum-ready solvers. 
Let $x \in \{0,1\}^n$ remain binary. Discretize the scalar $\phi$ on an interval $[\phi_{\min},\phi_{\max}]$ with $T_\phi$ binary bits:
\begin{align}
\phi \approx \phi_{\min} + \sum_{k=1}^{T_\phi} w_k q_k, \quad q_k \in \{0,1\},
\end{align}
where weights $w_k$ are geometric (1,2,4,\dots) scaled to the interval. We set $\phi_{\max}=\min\{\phi_{\max}^{\text{relax}},\;\min_i \alpha_i - \sum_j \max(\beta_{i,j},0)\}$ when optimality cuts exist (otherwise $\phi_{\max}=\phi_{\max}^{\text{relax}}$), and pick $\phi_{\min}$ a bit below that envelope to ensure dynamic range.

Each linear inequality is converted into a \emph{penalized equality} by introducing nonnegative slack variables that are themselves binary-encoded:
\begin{align}
Bx + s_B &= b', \\
(\phi-\phi_{\min}) + \beta_i^\top x + t_i &= \alpha_i - \phi_{\min}, \quad (\alpha_i,\beta_i) \in \mathcal{O}, \\
\gamma_j^\top x + u_j &= \rho_j, \quad (\rho_j,\gamma_j) \in \mathcal{F},
\end{align}
with $s_B, t_i, u_j \ge 0$ represented via $T_{\text{slack}}$ bits each. 
The resulting unconstrained objective is
\begin{align}
\min_{z \in \{0,1\}^N}\;& -\big(c^\top x + \phi\big) 
+ \lambda_B\,\|Bx + s_B - b'\|_2^2 
+ \lambda_O \sum_i \big((\phi-\phi_{\min}) + \beta_i^\top x + t_i - (\alpha_i - \phi_{\min})\big)^2 \\
&\quad + \lambda_F \sum_j \big(\gamma_j^\top x + u_j - \rho_j\big)^2,
\end{align}
where $z$ stacks all binary variables $[x;\,q;\,\text{slack bits}]$ and $(\lambda_B,\lambda_O,\lambda_F)$ are penalties chosen proportional to the problem scale (defaulting to $10\times$ a base magnitude derived from $\|c\|_1$ and the $\phi$ range). 
Because $z_i^2=z_i$ for binary variables, all linear and quadratic terms assemble into a symmetric QUBO matrix $Q$ with energy $E(z)=z^\top Q z$.


\subsection{Hybrid Quantum–Classical Solution of the QUBO}
\label{subsec:hybrid-solver}

We solve the master-problem QUBO from Sec.~\ref{subsec:mp_to_qubo} with a hybrid subQUBO pipeline that alternates quantum subproblem solves with lightweight classical refinement~\cite{boost2017partitioning,atobe2021hybrid,zhao2025clustering}. Let $E(z)=z^\top Q z$ be the QUBO energy with $z\in\{0,1\}^N$ (linear terms absorbed into $Q$). The procedure maintains an incumbent $z$ and repeatedly selects a subset $S\subset\{1,\dots,N\}$, builds a clamped subQUBO over $S$ by fixing $z_{\bar S}$, optimizes that subproblem on a quantum routine (e.g., QA or QAOA), merges the result back into the full vector, and then performs a short classical local-improvement sweep. Variables whose single-bit flips have similar impact are placed in the same group $S$, a conventional practice in subQUBO hybrids~\cite{boost2017partitioning}. Impacts are recomputed as the iterate changes, and the process repeats until no improvement is observed or a preset iteration budget is reached.

Given a current iterate $z$, the one-flip energy change used to score variable $i$ is
\begin{equation}
\label{eq:deltaE}
\Delta E_i \;=\; (1-2 z_i)\!\left(Q_{ii} + 2 \sum_{j\neq i} Q_{ij} z_j\right),
\end{equation}
which follows from $E(z\oplus e_i)-E(z)$. A batch $S$ is formed and all variables in $\bar S$ are clamped to their current values. The resulting subQUBO over $x_S\in\{0,1\}^{|S|}$ has the objective
\begin{equation}
\label{eq:subqubo}
f_S(x_S) \;=\; \sum_{i\in S} \big(Q_{ii}+d_i\big)\,x_i \;+\; \sum_{\substack{i,j\in S\\ i<j}} Q_{ij}\,x_i x_j,
\end{equation}
with the clamping-induced linear shift
\begin{equation}
\label{eq:shift}
d_i \;=\; \sum_{j\notin S} (Q_{ij}+Q_{ji})\,z_j.
\end{equation}
A quantum subroutine is applied to \eqref{eq:subqubo} to produce $x_S^\star$, possibly using multiple shots and keeping the lowest-energy sample. The full assignment is updated by setting $z_S\leftarrow x_S^\star$ while keeping $z_{\bar S}$ fixed.

Immediately after each quantum update, a classical 1-flip descent is applied to ensure the iterate is at least a local minimum with respect to single-bit moves. Using \eqref{eq:deltaE}, bits with $\Delta E_i<0$ are flipped and the test is repeated until no such bit remains. This sweep costs $\mathcal{O}(N^2)$ in the dense case (or $\mathcal{O}(\mathrm{nnz}(Q))$ if updates are maintained incrementally) and stabilizes progress between quantum calls.

The outer loop iterates over batches $S$, updates impacts via \eqref{eq:deltaE}, optionally re-partitions the variable set, and retains the best energy found so far. Typical stopping rules are: no energy improvement for several full passes over the variable set, reaching a fixed number of quantum calls, or meeting a target energy. After convergence, the final binary vector $z^\star$ is decoded back into the master-problem variables $(x,\phi)$, giving an approximate solution to the MILP master problem.

\section{Experimental Evaluation and Results Analysis}
\label{sec:results}

\subsection{Experimental Data Sources}
\label{subsec:data}

To evaluate the performance of the proposed Benders decomposition algorithm for crude oil scheduling, we conduct comprehensive computational experiments across a set of 15 test instances (Case 1–Case 15) of varying scale and complexity. The problem encompasses four key entities: vessels, storage tanks, blend tanks, and crude distillation units (CDUs), modeling the end-to-end supply chain from marine delivery to final distillation.

The instances are constructed following a progressive scaling strategy. As summarized in Table \ref{tab:problem_instances}, the number of vessels increases from 2 to 13, storage tanks from 2 to 6, blend tanks from 2 to 11, CDUs from 2 to 17, and scheduling periods from 3 to 50. This progression leads to exponential growth in model size: the number of discrete variables ranges from 50 to 5,350, continuous variables from 60 to 16,550, and constraints from 281 to 51,931. This multi-scale design enables a systematic assessment of the algorithm's scalability, convergence behavior, and solution quality.

All instance parameters—such as initial inventory, capacity limits, flow bounds, and time windows—are calibrated using real-world refinery data to ensure practical relevance. Among these, Case 15 is derived from a real-world crude oil scheduling scenario. It represents the largest and most complex instance, involving 13 vessels, 6 storage tanks, 11 blend tanks, 17 CDUs, 50 time intervals, and 212 pipeline connections. The resulting model contains 5,350 discrete variables and 51,931 constraints. In this highly complex case, traditional optimization methods tend to encounter difficulties in navigating the vast solution space and often converge to local optima. The reformulation of the master problem as a quadratic unconstrained binary optimization (QUBO) model facilitates the application of quantum-inspired or quantum algorithms, which demonstrate a stronger potential for escaping local optima and approaching a global optimum. The remaining instances are generated via simulation based on typical operational scenarios, collectively forming a robust and practical testbed.

\begin{figure}[htbp]
    \centering
    \includegraphics[width=0.9\columnwidth]{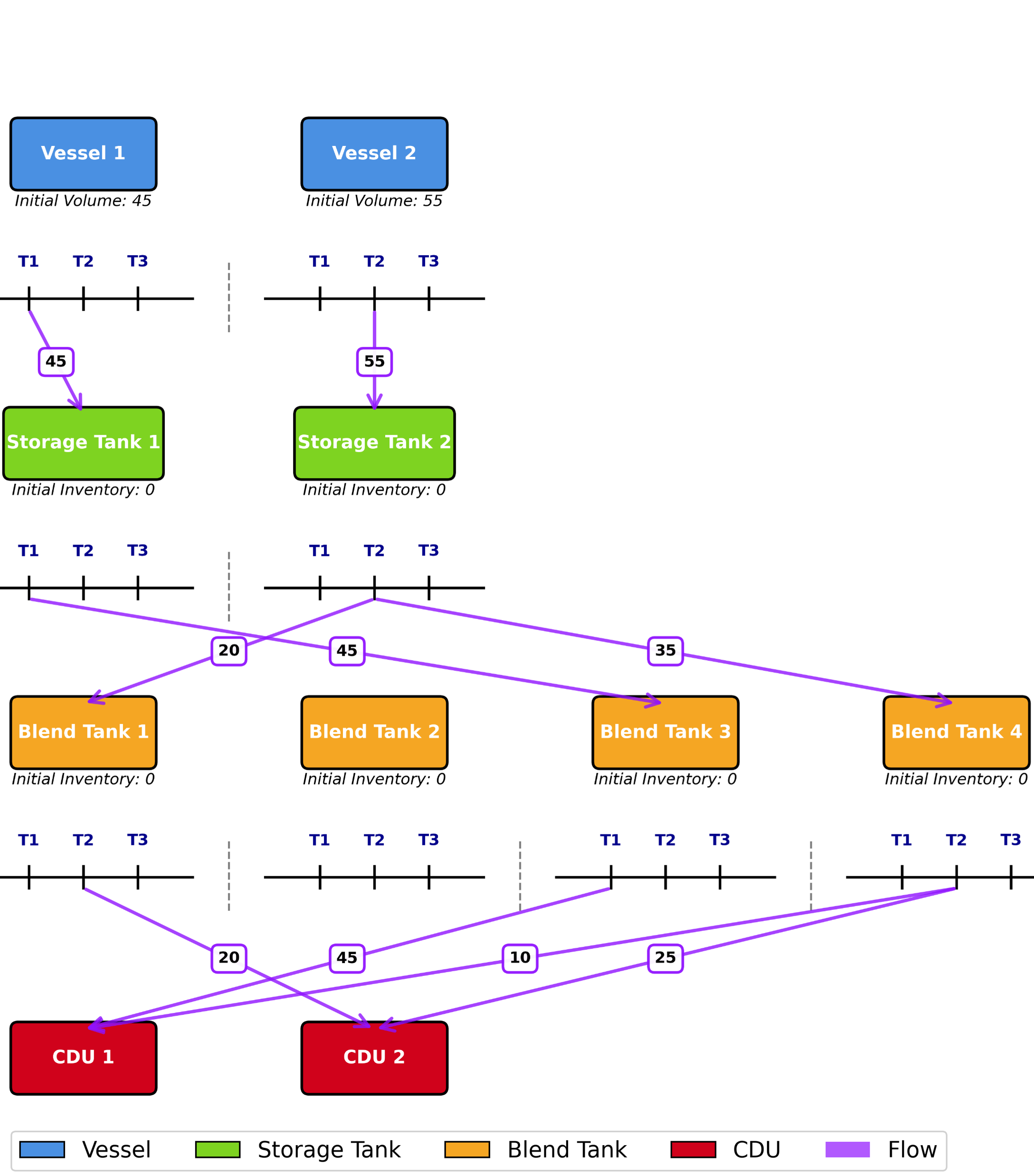}
    \caption{A schematic diagram of petroleum flow direction in the process of solving crude oil scheduling using benders}
    \label{fig:3}
\end{figure}

The analysis focuses on the convergence behavior of the Benders master-subproblem iterations and the computational impact of the QUBO reformulation. Algorithm performance is evaluated using objective value, computation time, and iteration count, providing insights into its robustness and efficiency. Figure \ref{fig:3} illustrates a schematic diagram of the oil flow direction within a Benders solution for the crude oil scheduling process.

\begin{table*}[htbp]
\centering
\caption{Problem Instance Parameter Statistics Table}
\small
\setlength{\tabcolsep}{1pt}  
\begin{tabular}{@{}clrrrrrrrrrr@{}}
\toprule
No. & Case & 
\shortstack{Number\\of\\Vessels} & 
\shortstack{Storage\\Tanks} & 
\shortstack{Feed\\Tanks} & 
\shortstack{Distillation\\Units} & 
\shortstack{Total\\Time\\Intervals} & 
\shortstack{Number\\of\\Pipelines} & 
\shortstack{Original\\Discrete\\Variables} & 
\shortstack{Original\\Continuous\\Variables} & 
\shortstack{Original\\Constraints} & 
\shortstack{Time\\Discrete\\Variables} \\
\midrule
1 & case1 & 2 & 2 & 2 & 2 & 5 & 12 & 50 & 60 & 281 & 50 \\
2 & case2 & 2 & 2 & 2 & 5 & 10 & 18 & 160 & 180 & 716 & 160 \\
3 & case3 & 2 & 2 & 4 & 2 & 3 & 20 & 42 & 60 & 251 & 42 \\
4 & case4 & 2 & 2 & 2 & 2 & 4 & 12 & 40 & 48 & 229 & 40 \\
5 & case5 & 6 & 2 & 2 & 2 & 9 & 20 & 198 & 180 & 906 & 198 \\
6 & case6 & 6 & 2 & 2 & 2 & 9 & 20 & 198 & 180 & 902 & 198 \\
7 & case7 & 6 & 2 & 2 & 2 & 15 & 20 & 330 & 300 & 1522 & 330 \\
8 & case8 & 6 & 2 & 2 & 2 & 20 & 20 & 440 & 400 & 2021 & 440 \\
9 & case9 & 6 & 6 & 2 & 2 & 30 & 52 & 660 & 1560 & 5897 & 660 \\
10 & case10 & 10 & 6 & 2 & 2 & 30 & 76 & 1020 & 2280 & 8725 & 1020 \\
11 & case11 & 10 & 6 & 4 & 2 & 30 & 92 & 1140 & 2760 & 9929 & 1140 \\
12 & case12 & 10 & 6 & 4 & 2 & 30 & 92 & 1140 & 2760 & 9929 & 1140 \\
13 & case13 & 10 & 6 & 4 & 5 & 30 & 104 & 1500 & 3120 & 11012 & 1500 \\
14 & case14 & 10 & 6 & 4 & 17 & 30 & 152 & 2940 & 4560 & 15344 & 2940 \\
15 & case15 & 13 & 6 & 11 & 17 & 50 & 212 & 5350 & 16550 & 51931 & 5350 \\
\bottomrule
\end{tabular}
\label{tab:problem_instances}
\end{table*}

To analyze the impact of different algorithms on problem scale, Table.\ref{tab:discrete_vars_compact} compares key metrics across variable and constraint dimensions. The results demonstrate that the Benders decomposition framework largely preserves the original problem structure: the number of discrete variables remains identical to the original formulation; continuous variables increase by only one auxiliary variable ($\phi$) for objective value transmission; constraint counts are maintained in the master problem, while subproblems handle only continuous-related constraints, progressively approaching optimality through iteratively generated Benders cuts. This design ensures controllable problem size growth and enables direct utilization of the original problem's structural information. In contrast, alternative approaches (e.g., direct Gurobi solving, spectral clustering, and various heuristic methods) typically require more complex model reformulations or variable encodings, leading to significant changes in variable and constraint counts and increasing the explicit complexity of the problem.

\begin{table}[htbp]
\centering
\caption{Comparison of Discrete Variable Counts Across Algorithmic Formulations}
\small  
\setlength{\tabcolsep}{2pt}  
\begin{tabular}{@{}lrrrrrrrrr@{}}
\toprule
Case & Orig. & 
\shortstack{Benders\\(Quantum)} &  
\shortstack{Gurobi\\(Quantum)} &  
\shortstack{Spectral Clust.\\(Quantum)} &  
\shortstack{Cluster\\(Classical)} &  
\shortstack{Genetic\\(Classical)} &  
\shortstack{Impact\\(Classical)} &  
\shortstack{Tabu\\(Classical)} \\  
\midrule
case1 & 50 & 50 & 70 & 90 & 30 & 30 & 30 & 30 \\
case2 & 160 & 160 & 260 & 360 & 60 & 60 & 60 & 60 \\
case3 & 42 & 42 & 114 & 138 & 18 & 18 & 18 & 18 \\
case4 & 40 & 40 & 56 & 72 & 24 & 24 & 24 & 24 \\
case5 & 198 & 198 & 234 & 270 & 162 & 162 & 162 & 162 \\
case6 & 198 & 198 & 234 & 270 & 162 & 162 & 162 & 162 \\
case7 & 330 & 330 & 390 & 450 & 270 & 270 & 270 & 270 \\
case8 & 440 & 440 & 520 & 600 & 360 & 360 & 360 & 360 \\
case9 & 660 & 660 & 780 & 900 & 540 & 540 & 540 & 540 \\
case10 & 1020 & 1020 & 1140 & 1260 & 900 & 900 & 900 & 900 \\
case11 & 1140 & 1140 & 1860 & 2100 & 900 & 900 & 900 & 900 \\
case12 & 1140 & 1140 & 1860 & 2100 & 900 & 900 & 900 & 900 \\
case13 & 1500 & 1500 & 3300 & 3900 & 900 & 900 & 900 & 900 \\
case14 & 2940 & 2940 & 9060 & 11100 & 900 & 900 & 900 & 900 \\
case15 & 5350 & 5350 & 104800 & 114150 & 1950 & 1950 & 1950 & 1950 \\
\bottomrule
\end{tabular}
\label{tab:discrete_vars_compact}
\end{table}

\begin{table}[htbp]
\centering
\caption{Comparison of Continuous Variable Counts Across Algorithmic Formulations}
\small  
\setlength{\tabcolsep}{2pt}  
\begin{tabular}{@{}lrrrrrrrrr@{}}
\toprule
Case & Orig. & 
\shortstack{Benders\\(Quantum)} &  
\shortstack{Gurobi\\(Quantum)} &  
\shortstack{Spectral Clust.\\(Quantum)} &  
\shortstack{Cluster\\(Classical)} &  
\shortstack{Genetic\\(Classical)} &  
\shortstack{Impact\\(Classical)} &  
\shortstack{Tabu\\(Classical)} \\  
\midrule
case1 & 60 & 61 & 86 & 148 & 80 & 80 & 80 & 80 \\
case2 & 180 & 181 & 229 & 411 & 220 & 220 & 220 & 220 \\
case3 & 60 & 61 & 86 & 148 & 78 & 78 & 78 & 78 \\
case4 & 48 & 49 & 70 & 120 & 64 & 64 & 64 & 64 \\
case5 & 180 & 181 & 222 & 408 & 216 & 216 & 216 & 216 \\
case6 & 180 & 181 & 222 & 408 & 216 & 216 & 216 & 216 \\
case7 & 300 & 301 & 366 & 672 & 360 & 360 & 360 & 360 \\
case8 & 400 & 401 & 486 & 892 & 480 & 480 & 480 & 480 \\
case9 & 1560 & 1561 & 1810 & 3376 & 1800 & 1800 & 1800 & 1800 \\
case10 & 2280 & 2281 & 2530 & 4820 & 2520 & 2520 & 2520 & 2520 \\
case11 & 2760 & 2761 & 3072 & 5842 & 3060 & 3060 & 3060 & 3060 \\
case12 & 2760 & 2761 & 3072 & 5842 & 3060 & 3060 & 3060 & 3060 \\
case13 & 3120 & 3121 & 3435 & 6565 & 3420 & 3420 & 3420 & 3420 \\
case14 & 4560 & 4561 & 4887 & 9457 & 4860 & 4860 & 4860 & 4860 \\
case15 & 16550 & 16551 & 17434 & 33997 & 17400 & 17400 & 17400 & 17400 \\
\bottomrule
\end{tabular}
\label{tab:continuous_vars_compact}
\end{table}

\begin{table}[htbp]
\centering
\caption{Comparison of Constraint Counts Across Algorithmic Formulations}
\small  
\setlength{\tabcolsep}{2pt}  
\begin{tabular}{@{}lrrrrrrrrr@{}}
\toprule
Case & Orig. & 
\shortstack{Benders\\(Quantum)} &  
\shortstack{Gurobi\\(Quantum)} &  
\shortstack{Spectral Clust.\\(Quantum)} &  
\shortstack{Cluster\\(Classical)} &  
\shortstack{Genetic\\(Classical)} &  
\shortstack{Impact\\(Classical)} &  
\shortstack{Tabu\\(Classical)} \\  
\midrule
case1 & 281 & 281 & 14 & 188 & 137 & 137 & 137 & 137 \\
case2 & 716 & 716 & 24 & 605 & 325 & 325 & 325 & 325 \\
case3 & 251 & 251 & 10 & 233 & 119 & 119 & 119 & 119 \\
case4 & 229 & 229 & 12 & 153 & 112 & 112 & 112 & 112 \\
case5 & 906 & 906 & 30 & 520 & 465 & 465 & 465 & 465 \\
case6 & 902 & 902 & 30 & 516 & 465 & 465 & 465 & 465 \\
case7 & 1522 & 1522 & 42 & 871 & 759 & 759 & 759 & 759 \\
case8 & 2021 & 2021 & 52 & 1156 & 1004 & 1004 & 1004 & 1004 \\
case9 & 5897 & 5897 & 192 & 2809 & 2578 & 2578 & 2578 & 2578 \\
case10 & 8725 & 8725 & 200 & 3915 & 3790 & 3790 & 3790 & 3790 \\
case11 & 9929 & 9929 & 200 & 5297 & 4332 & 4332 & 4332 & 4332 \\
case12 & 9929 & 9929 & 200 & 5297 & 4332 & 4332 & 4332 & 4332 \\
case13 & 11012 & 11012 & 200 & 7460 & 4695 & 4695 & 4695 & 4695 \\
case14 & 15344 & 15344 & 200 & 16112 & 6147 & 6147 & 6147 & 6147 \\
case15 & 51931 & 51931 & 326 & 131884 & 20123 & 20123 & 20123 & 20123 \\
\bottomrule
\end{tabular}
\label{tab:constraints_compact}
\end{table}

\subsection{Experimental Results}
\label{subsec:results}

\begin{figure}[htbp]
    \centering
    \includegraphics[width=0.9\columnwidth]{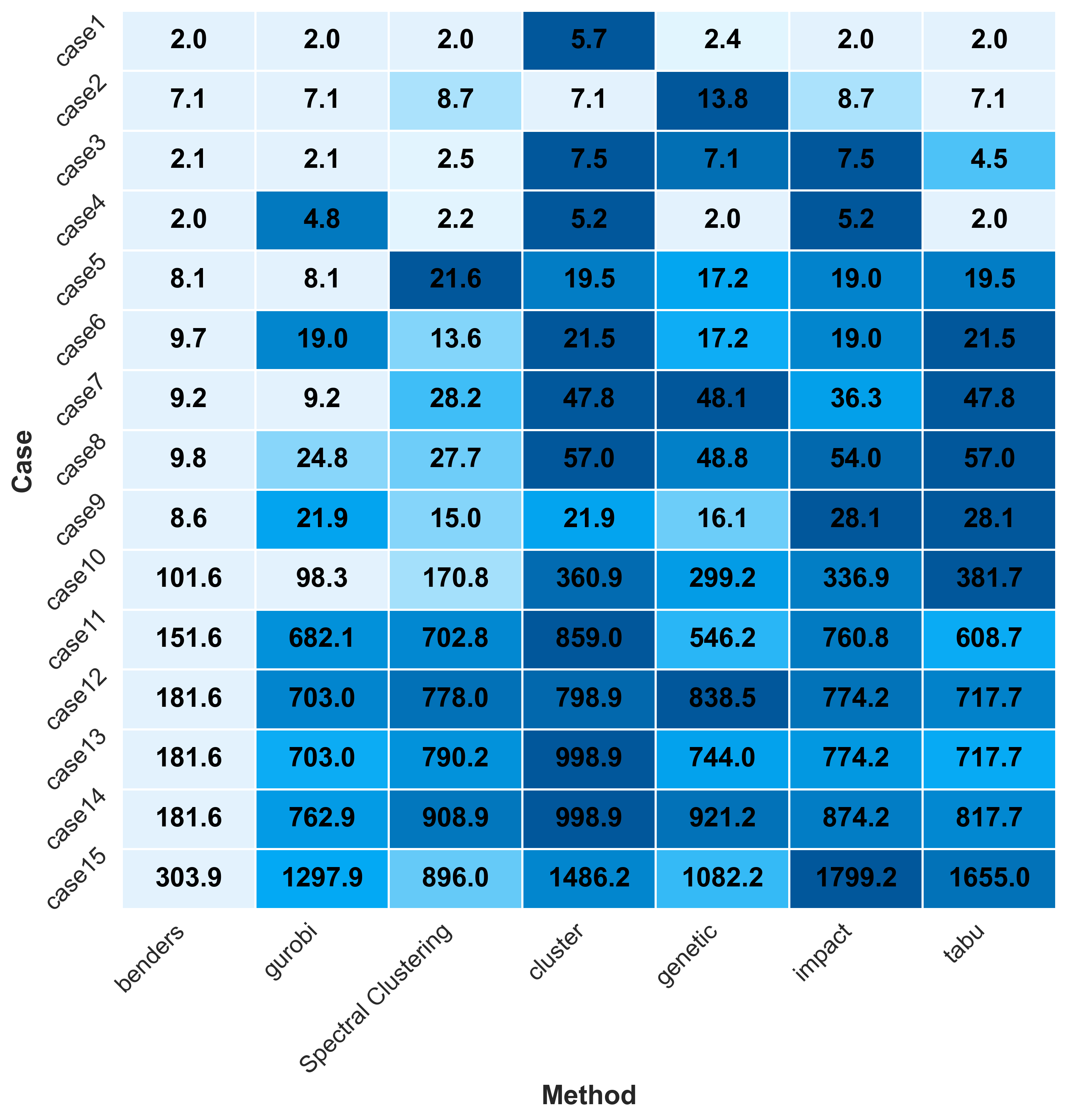}
    \caption{Target cost results of various algorithms}
    \label{fig:4a}
\end{figure}

\begin{figure}[htbp]
    \centering
    \includegraphics[width=0.9\columnwidth]{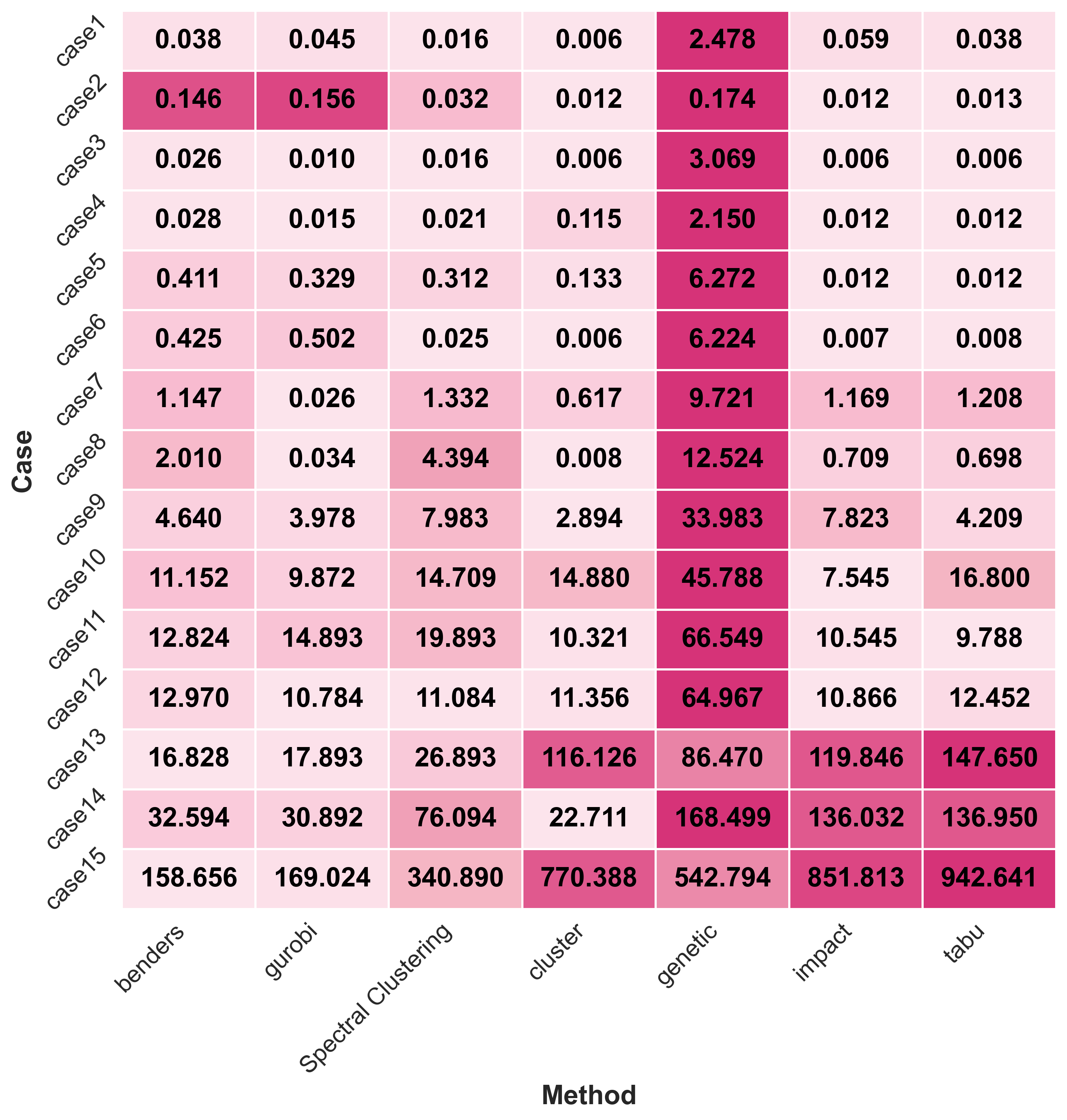}
    \caption{Calculation time of various algorithms}
    \label{fig:4b}
\end{figure}

The comprehensive experimental evaluation presented in Figure.\ref{fig:4a} and \ref{fig:4b} reveals significant performance disparities among the evaluated optimization approaches for crude oil scheduling problems. The Benders decomposition method demonstrates superior performance across both cost minimization and computational efficiency metrics, achieving an average operational cost of 77.35 units compared to 289.75-379.73 units for alternative methods, representing a cost reduction of approximately 73-80\%. This substantial cost advantage is particularly critical in crude oil scheduling operations, where even marginal improvements translate to substantial economic benefits given the high-value nature of petroleum products and the continuous operation of refineries. The computational efficiency of the Benders method is equally noteworthy, with an average execution time of 16.93 seconds, which is comparable to Gurobi (17.23 seconds) but significantly faster than metaheuristic approaches such as genetic algorithms (70.11 seconds) and tabu search (84.83 seconds), demonstrating a 76-80\% reduction in computational time.

The superior performance of the Benders decomposition approach can be attributed to its inherent ability to exploit the underlying structure of the crude oil scheduling problem, which typically exhibits natural decompositions between discrete assignment decisions and continuous flow optimization. Unlike metaheuristic methods that rely on stochastic search mechanisms and may converge to suboptimal solutions, the Benders method systematically explores the solution space through a cutting-plane approach, ensuring convergence to near-optimal solutions while maintaining computational tractability. The relatively low standard deviations observed in the Benders method's performance (98.30 for cost and 40.27 for time) further indicate its robustness and reliability across diverse problem instances, a crucial attribute for industrial applications where solution consistency is paramount. In contrast, methods such as spectral clustering and impact-based approaches exhibit higher variability, with standard deviations exceeding 500 for cost metrics, suggesting limited reliability for practical implementation in time-sensitive refinery operations.

From a practical perspective, the demonstrated advantages of the Benders decomposition method have profound implications for refinery scheduling operations. The ability to generate high-quality solutions within approximately 17 seconds enables real-time or near-real-time decision support, which is essential for responding to dynamic operational conditions such as crude oil arrival delays, equipment maintenance requirements, or unexpected demand fluctuations. The substantial cost savings achieved through this method not only improve operational profitability but also enhance the overall competitiveness of refinery operations in an increasingly challenging market environment. Furthermore, the method's consistent performance across multiple test cases suggests its generalizability to various refinery configurations and operational scenarios, making it a robust tool for strategic planning and operational optimization in the petroleum industry.

The comprehensive scoring analysis in Table~\ref{tab:algorithm_performance} evaluates algorithm performance on cost minimization and computational efficiency, normalized to a 0--100 scale. The Benders decomposition method achieves optimal performance with perfect scores in both metrics. Gurobi shows high computational efficiency but moderate cost performance. Spectral clustering offers balanced results, while metaheuristics (cluster, genetic, impact, tabu) perform lower due to limited cost-effectiveness and speed.

Normalization formulas:
\begin{itemize}
    \item \textbf{Cost Score:} 
    \[
    \text{Cost\_Score} = 100 \times \frac{\text{Cost}_{\max} - \text{Cost}}{\text{Cost}_{\max} - \text{Cost}_{\min}}
    \]
    where \(\text{Cost}_{\min}\) corresponds to 100 and \(\text{Cost}_{\max}\) to 0.
    \item \textbf{Time Score:}
    \[
    \text{Time\_Score} = 100 \times \frac{\text{Time}_{\max} - \text{Time}}{\text{Time}_{\max} - \text{Time}_{\min}}
    \]
    where \(\text{Time}_{\min}\) corresponds to 100 and \(\text{Time}_{\max}\) to 0.
\end{itemize}
The total score is computed as:
\[
\text{Total\_Score} =   \text{Cost\_Score}  + \text{Time\_Score}
\]
with weights of 0.6 for cost and 0.4 for time.

\begin{table}[htbp]
\centering
\caption{Algorithm Performance Comparison}
\small
\setlength{\tabcolsep}{1.5pt}
\begin{tabular}{@{}lrrrrrrr@{}}
\toprule
Metric & Benders & Gurobi & Spectral Clustering & Cluster & Genetic & Impact & Tabu \\
\midrule
Cost Score & 100.00 & 29.76 & 29.27 & 0.00 & 24.07 & 4.33 & 13.40 \\
Time Score & 100.00 & 99.55 & 75.48 & 31.70 & 21.68 & 12.37 & 0.00 \\
Total Score & 200.00 & 129.31 & 104.75 & 31.70 & 45.75 & 16.71 & 13.40 \\
\bottomrule
\end{tabular}
\label{tab:algorithm_performance}
\end{table}

\begin{figure}[htbp]
    \centering
    \includegraphics[width=0.9\columnwidth]{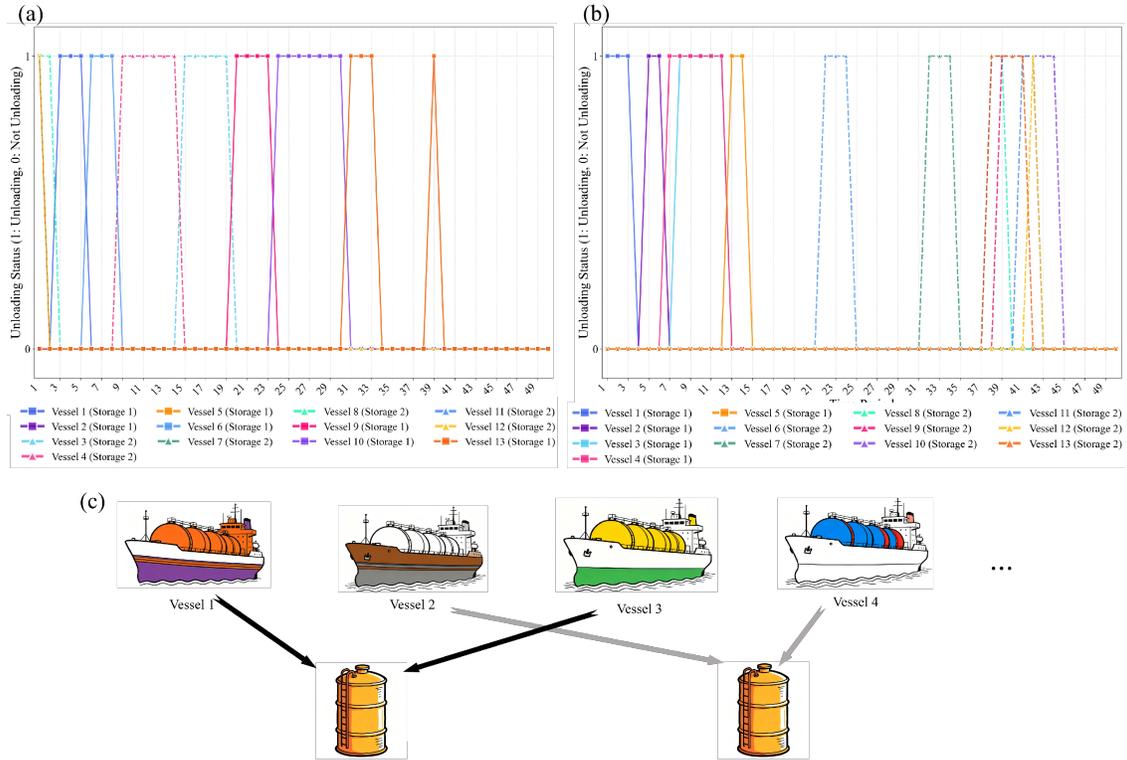}
    \caption{Traditional algorithms are prone to getting stuck in local optima in special situations}
    \label{fig:5}
\end{figure}

Figure.\ref{fig:5} provides a direct comparison of the scheduling results obtained from Benders decomposition and a Genetic Algorithm (GA). The schedule generated by the Benders method (Fig. 5a) exhibits a balanced distribution of tankers between Storage Tanks 1 and 2, a direct outcome of its iterative master-subproblem framework which enables global resource coordination to prevent capacity violations. In stark contrast, the GA's schedule (Fig. 5b) is characterized by a suboptimal pattern: driven by a greedy fitness function, approximately 35\% of tankers are densely clustered in the lower-cost Tank 1 during early periods (T1--T12), forcing a costly late-stage shift of the remaining 45\% of tankers to Tank 2 (T32--T45) after Tank 1's capacity is exhausted.

The ability of Benders decomposition to avoid such local optima is rooted in its integrated global feedback mechanism. The Master Problem (MP) handles the discrete assignment decisions, while the Subproblem (SP) evaluates the resulting continuous flows and physical constraints. Critically, when the SP detects that a decision from the MP would lead to an issue like tank overflow, it communicates this global consequence back to the MP via an \textbf{optimality cut}, formulated as \(\phi \leq \mu^T (b - A x)\), where \(\mu\) represents the dual variables. This feedback allows the MP to explicitly foresee and proactively avoid resource conflicts in subsequent iterations, leading to a strategically balanced allocation.

This comparison underscores a fundamental distinction between mathematical decomposition and traditional heuristics. By leveraging optimality cuts, Benders decomposition maintains a global perspective that identifies and avoids locally attractive but systemically costly decisions. Methods like the GA, which lack such a mechanism and evaluate decisions based primarily on immediate fitness, are consequently prone to becoming trapped in inferior local solutions, resulting in significantly higher overall costs..

\section{Conclusion}
\label{sec:conclusion}

This study presents a novel hybrid quantum-classical optimization framework designed to address the computationally intractable crude oil scheduling problem in refinery operations. By employing Benders decomposition, the complex Mixed-Integer Linear Programming (MILP) model is systematically decoupled into a discrete Master Problem (MP) and a continuous Subproblem (SP). Formulating the MP as a Quadratic Unconstrained Binary Optimization (QUBO) model allows the framework to exploit the expansive search capabilities of quantum and hybrid solvers. Concurrently, the SP ensures strict adherence to mass balance, pipeline network constraints, and quality specifications by generating iterative optimality and feasibility cuts.

Extensive computational experiments across 15 multi-scale instances systematically validate the efficacy and scalability of the proposed framework. By strictly enforcing complex physical constraints through iterative Benders cuts, the approach successfully avoids the severe penalty costs and local optima that typically compromise traditional metaheuristics (e.g., Genetic Algorithms and Tabu Search). Crucially, the framework achieves high-quality schedules with an average computational time of approximately 17 seconds, delivering solution optimality on par with state-of-the-art exact solvers like Gurobi. This combination of mathematical rigor and hybrid computational efficiency ensures robust scalability, fulfilling the stringent requirements for near real-time industrial decision support.

A fundamental advantage of this architecture is its structural immunity to local optima, a severe limitation in greedy heuristic methods. By translating global system constraints into explicit dual feedback, the algorithm anticipates resource bottlenecks and ensures a strategically balanced allocation of vessels and tanks. Ultimately, this research bridges the gap between emerging quantum optimization techniques and complex process systems engineering, providing a highly scalable, robust, and economically impactful solution for modern petroleum supply chains.

\begin{acknowledgement}
This work was supported by Beijing Municipal Science \& Technology Commission, Beijing Science and Technology Plan ($Z241100004224034$).
\end{acknowledgement}

\begin{suppinfo}

\section{Problem Overview}
\label{app:prob_overview}
This appendix presents the detailed Mixed-Integer Linear Programming (MILP) formulation for the refinery front-end scheduling problem. It defines the nomenclature, decision variables, objective function, and the complete set of constraints governing the system.

The physical topology of the system is modeled as a multi-stage network comprising four primary entity levels:
\begin{enumerate}
    \item \textbf{Vessels ($V$)}: The material sources of the system. Each vessel carries specific types of crude oil and arrives at the port within a predetermined time window. Due to the scarcity of berthing resources, vessels may need to wait at the anchorage. If the retention time exceeds the allowable laytime, the refinery incurs high demurrage costs.
    \item \textbf{Storage Tanks ($I$)}: Directly connected to the unloading pipelines at the dock, these tanks receive crude oil unloaded from vessels, acting as the primary buffer pool for the system.
    \item \textbf{Charging/Blending Tanks ($J$)}: Located downstream of the storage tanks, these tanks receive crude oil for necessary blending or settling before feeding the production units. They are critical nodes connecting inventory to production.
    \item \textbf{Crude Distillation Units ($L$)}: The material sinks of the system. CDUs must maintain a continuous and stable feed rate to meet production demand ($DM$). Any interruption in supply leads to significant economic losses and safety risks.
\end{enumerate}

Within the scheduling horizon $T$, decision-makers must formulate a precise logistics plan that satisfies all physical connectivity constraints, tank capacity limits, flow rate restrictions, and production demands. Mathematically, this problem is characterized as an optimization problem involving complex intermediate constraints and mixed variables.

The relevant notation and parameters are defined in Table \ref{tab:params}.

\begin{longtable}{l l p{7cm} l}
\caption{Nomenclature and Parameters} \label{tab:params} \\
\toprule
\textbf{Category} & \textbf{Symbol} & \textbf{Description} & \textbf{Unit/Note} \\
\midrule
\endfirsthead
\toprule
\textbf{Category} & \textbf{Symbol} & \textbf{Description} & \textbf{Unit/Note} \\
\midrule
\endhead
\bottomrule
\endfoot

\textbf{Sets} & $V$ & Set of crude oil vessels, indexed by $v$ & $\{1, \dots, |V|\}$ \\
 & $I$ & Set of primary storage tanks, indexed by $i$ & $\{1, \dots, |I|\}$ \\
 & $J$ & Set of charging/blending tanks, indexed by $j$ & $\{1, \dots, |J|\}$ \\
 & $L$ & Set of Crude Distillation Units (CDUs), indexed by $l$ & $\{1, \dots, |L|\}$ \\
 & $T$ & Set of discrete time periods, indexed by $t$ & $\{1, \dots, |T|\}$ \\
\midrule
\textbf{Time} & $T^{arr}_v$ & Earliest arrival time for vessel $v$ & Time step \\
 & $T^{lea}_v$ & Latest departure time for vessel $v$ & Time step \\
 & $D_v$ & Duration required for vessel $v$ to complete unloading & Time steps \\
\midrule
\textbf{Cost} & $C^{sea}_v$ & Demurrage rate (waiting cost) for vessel $v$ & Currency/step \\
 & $C^{unload}_v$ & Fixed operation cost for unloading vessel $v$ & Currency/ops \\
 & $C^{inv}_i, C^{inv}_j$ & Unit inventory holding cost for tank $i$ or $j$ & Currency/(ton$\cdot$step) \\
 & $C^{setup}_l$ & Setup/Switching cost for feeding CDU $l$ & Currency/ops \\
\midrule
\textbf{Capacity} & $V^{ini}_v$ & Initial crude volume on vessel $v$ & Tons \\
\textbf{\& Flow} & $V^{ini}_i, V^{ini}_j$ & Initial inventory level in tanks $i, j$ & Tons \\
 & $V^{min/max}_{i/j}$ & Safety inventory lower/upper bounds for tanks $i, j$ & Tons \\
 & $F^{min/max}_{vs}$ & Flow bounds from vessel $v$ to tank $i$ & Tons/step \\
 & $F^{min/max}_{bs}$ & Flow bounds from tank $i$ to charging tank $j$ & Tons/step \\
 & $F^{min/max}_{bc}$ & Flow bounds from charging tank $j$ to CDU $l$ & Tons/step \\
 & $DM_l$ & Total crude demand for CDU $l$ over the horizon & Tons \\
\midrule
\textbf{Topology} & $A_{vi}$ & Binary parameter, 1 if vessel $v$ can unload to tank $i$ & 0-1 Matrix \\
 & $A_{ij}$ & Binary parameter, 1 if tank $i$ can transfer to tank $j$ & 0-1 Matrix \\
 & $A_{jl}$ & Binary parameter, 1 if tank $j$ can feed CDU $l$ & 0-1 Matrix \\
\end{longtable}

\subsection{Decision Variables}
\label{app:variables}
To achieve the optimization goals of the refinery front-end scheduling, we must formulate two layers of coupled plans: high-level discrete event scheduling and low-level continuous material transport.

First, \textbf{Discrete Scheduling Planning} addresses the assignment of "time" and "states". We must determine the specific unloading time windows for each vessel, including when to berth and start unloading and when to finish the operation. Additionally, within the complex pipeline network, the physical connection status between charging tanks $j$ and CDUs $l$ must be dynamically determined to avoid operational costs associated with frequent pipeline switching.

Second, \textbf{Continuous Logistics Planning} addresses the allocation of "quantities" and "routes". Given the determined time windows and connection states, we must precisely calculate the crude oil flow rates through each node at every time step. This includes the unloading rate from vessels to primary tanks, the transfer rate from primary to secondary tanks, and the feed rate into the CDUs.

It is worth noting that in the mathematical formulation of this model, the inventory levels of storage nodes are not treated as independent decision variables. Instead, they are state quantities derived from cumulative flows via the law of conservation of mass.

The core decision variables involved in the model are defined in Table \ref{tab:vars}.

\begin{table}[ht]
\centering
\caption{Definition of Decision Variables}
\label{tab:vars}
\begin{tabular}{l l l p{8cm}}
\toprule
\textbf{Category} & \textbf{Symbol} & \textbf{Domain} & \textbf{Description} \\
\midrule
\textbf{Binary} & $x^{W}_{v,t}$ & $\{0, 1\}$ & \textbf{Active State}: 1 if vessel $v$ is berthed or unloading at time $t$; 0 otherwise. \\
\textbf{Variables} & $x^{S}_{v,t}$ & $\{0, 1\}$ & \textbf{Start Unloading}: 1 if vessel $v$ \textbf{starts} unloading at time $t$; 0 otherwise. \\
 & $x^{E}_{v,t}$ & $\{0, 1\}$ & \textbf{End Unloading}: 1 if vessel $v$ \textbf{finishes} unloading at time $t$; 0 otherwise. \\
 & $z_{j,l,t}$ & $\{0, 1\}$ & \textbf{Pipeline Connection}: 1 if the pipeline between tank $j$ and CDU $l$ is active at time $t$; 0 otherwise. \\
\midrule
\textbf{Continuous} & $f^{VS}_{v,i,t}$ & $\mathbb{R}_{\ge 0}$ & \textbf{Unloading Flow}: Volume/mass transferred from vessel $v$ to tank $i$ at time $t$. \\
\textbf{Variables} & $f^{SB}_{i,j,t}$ & $\mathbb{R}_{\ge 0}$ & \textbf{Transfer Flow}: Volume/mass transferred from tank $i$ to charging tank $j$ at time $t$. \\
 & $f^{BC}_{j,l,t}$ & $\mathbb{R}_{\ge 0}$ & \textbf{Feed Flow}: Volume/mass fed from charging tank $j$ to CDU $l$ at time $t$. \\
\bottomrule
\end{tabular}
\end{table}

\subsection{Objective Function}
\label{app:obj_func}
The core objective of the schedule is to minimize the Total Operating Cost of the crude oil journey from the dock to the processing units. Mathematically, we construct the objective function as a linear maximization problem (maximizing negative cost), i.e., $\max (-Z)$, where $Z$ is the total cost.

The total cost function consists of four components: demurrage costs, fixed unloading costs, pipeline setup costs, and inventory holding costs. Based on the defined decision variables, the objective function expression is as follows:

\begin{equation} \label{eq:obj_total}
\text{Minimize } Z = Z_{dem} + Z_{fix} + Z_{setup} + Z_{hold}
\end{equation}

The mathematical expression and physical interpretation of each cost term are detailed below:

\paragraph{(1) Demurrage and Fixed Unloading Costs ($Z_{dem} + Z_{fix}$)}
This component corresponds to the coefficient vector $c$ for binary variables in the model. Demurrage depends on the delay of the vessel's actual start time relative to its arrival time, while the fixed cost is a one-time fee.

\begin{equation} \label{eq:obj_dem_fix}
Z_{dem} + Z_{fix} = \sum_{v \in V} \sum_{t \in T} \left[ C^{sea}_v \cdot \max(0, t - T^{arr}_v) + C^{unload}_v \right] \cdot x^S_{v,t}
\end{equation}

Here, $x^S_{v,t}$ is the binary variable for "starting" unloading. If vessel $v$ starts unloading at time $t$, a fixed cost $C^{unload}_v$ is incurred immediately, and the cumulative demurrage is calculated based on the delay $(t - T^{arr}_v)$.

\paragraph{(2) Pipeline Setup Costs ($Z_{setup}$)}
This component penalizes frequent pipeline switching operations, corresponding to the cost coefficients for binary variable $z$.

\begin{equation} \label{eq:obj_setup}
Z_{setup} = \sum_{j \in J} \sum_{l \in L} \sum_{t \in T} C^{setup}_l \cdot z_{j,l,t}
\end{equation}

Whenever the connection between charging tank $j$ and CDU $l$ is active, a corresponding maintenance or operational fee is generated.

\paragraph{(3) Cumulative Inventory Holding Costs ($Z_{hold}$)}
This component corresponds to the coefficient vector $h$ for continuous variables. Since this model eliminates explicit inventory state variables via cumulative flow constraints, the inventory holding cost is rewritten as a function of flow rates. Based on the principle of integration, a unit of crude oil entering a tank at time $t$ will generate holding costs for the remaining period $[t, |T|]$ if it does not flow out.

Thus, the inventory cost is decomposed into the net present value of inflows and outflows:

\begin{equation} \label{eq:obj_hold}
\begin{aligned}
Z_{hold} = & \sum_{t \in T} ( |T| - t ) \cdot \Bigg[ \sum_{v \in V}\sum_{i \in I} C^{inv}_i \cdot f^{VS}_{v,i,t} \\
& + \sum_{i \in I}\sum_{j \in J} (C^{inv}_j - C^{inv}_i) \cdot f^{SB}_{i,j,t} \\
& - \sum_{j \in J}\sum_{l \in L} C^{inv}_j \cdot f^{BC}_{j,l,t} \Bigg]
\end{aligned}
\end{equation}

\begin{itemize}
    \item \textbf{First term ($f^{VS}$)}: Crude oil entering storage tank $i$ from a vessel increases inventory at level $I$, generating a positive holding cost $C^{inv}_i$.
    \item \textbf{Second term ($f^{SB}$)}: Crude oil transferred from storage tank $i$ to charging tank $j$ reduces inventory at level $I$ (cost saving $-C^{inv}_i$) while increasing inventory at level $J$ (cost $+C^{inv}_j$).
    \item \textbf{Third term ($f^{BC}$)}: Crude oil leaving charging tank $j$ to the CDU reduces inventory at level $J$, generating a negative holding cost (cost saving).
    \item \textbf{Coefficient $(|T| - t)$}: Represents the cumulative weight of the unit flow on inventory levels for all future time steps.
\end{itemize}

\subsection{Constraints}
\label{app:constraints}
The constraints aim to capture the physical limitations, operational rules, and supply-demand balance of the refinery front-end scheduling. The constraint set $\mathcal{C}$ can be divided into three categories: Scheduling Logic, Cumulative Inventory Balance, and Operational Constraints.

\paragraph{(1) Scheduling Logic \& Time Windows}
These constraints involve only binary variables and ensure the logical consistency of discrete events.

\begin{itemize}
    \item \textbf{Task Uniqueness and Sequence}: Each vessel $v$ must start unloading exactly once and finish unloading exactly once within the entire scheduling horizon.
    \begin{equation} \label{eq:con_unique}
    \sum_{t \in T} x^S_{v,t} = 1, \quad \sum_{t \in T} x^E_{v,t} = 1 \quad \forall v \in V
    \end{equation}

    \item \textbf{State Coupling}: The "active" state $x^W_{v,t}$ is strictly bounded by "start" and "end" events. $x^W_{v,t}$ can only be active after the start time and before the end time.
    \begin{equation} \label{eq:con_state}
    x^W_{v,t} \le \sum_{\tau=1}^t x^S_{v,\tau}, \quad x^W_{v,t} \le 1 - \sum_{\tau=1}^{t-1} x^E_{v,\tau} \quad \forall v, t
    \end{equation}

    \item \textbf{Duration Constraint}: Once a vessel starts unloading, it must operate continuously for $D_v$ time steps.
    \begin{equation} \label{eq:con_duration}
    \sum_{\tau=t}^{t+D_v-1} x^W_{v,\tau} \ge D_v \cdot x^S_{v,t} \quad \forall v, t
    \end{equation}

    \item \textbf{Berth Resource Limit}: Assuming a single berth (or limited capacity), the number of vessels in active state at any time cannot exceed one.
    \begin{equation} \label{eq:con_berth}
    \sum_{v \in V} x^W_{v,t} \le 1 \quad \forall t \in T
    \end{equation}

    \item \textbf{Time Window Restrictions}: Vessels can only start after the allowable arrival time and must finish before the latest departure time.
    \begin{equation} \label{eq:con_tw}
    x^S_{v,t} = 0 \quad \text{if } t < T^{arr}_v, \quad x^E_{v,t} = 0 \quad \text{if } t > T^{lea}_v
    \end{equation}
\end{itemize}

\paragraph{(2) Cumulative Inventory \& Material Balance}
The model adopts an inventory-variable-free strategy, explicitly expressing material conservation by summing flow variables over time.

\begin{itemize}
    \item \textbf{Vessel Total Balance}: The total flow unloaded from a vessel must equal its initial cargo volume.
    \begin{equation} \label{eq:con_vessel_bal}
    \sum_{i \in I} \sum_{t \in T} f^{VS}_{v,i,t} = V^{ini}_v \quad \forall v \in V
    \end{equation}

    \item \textbf{Primary Tank Inventory Bounds}: For any time $t$, the current inventory of tank $i$ (Initial + Cumulative Inflow - Cumulative Outflow) must remain within the safety capacity bounds $[V^{min}_i, V^{max}_i]$.
    \begin{equation} \label{eq:con_tank_i}
    V^{min}_i \le V^{ini}_i + \sum_{\tau=1}^t \left( \sum_{v \in V} f^{VS}_{v,i,\tau} - \sum_{j \in J} f^{SB}_{i,j,\tau} \right) \le V^{max}_i \quad \forall i, t
    \end{equation}

    \item \textbf{Secondary Tank Inventory Bounds}: Similarly, the inventory level of charging tank $j$ is determined by inflows from tank $i$ and outflows to CDU $l$.
    \begin{equation} \label{eq:con_tank_j}
    V^{min}_j \le V^{ini}_j + \sum_{\tau=1}^t \left( \sum_{i \in I} f^{SB}_{i,j,\tau} - \sum_{l \in L} f^{BC}_{j,l,\tau} \right) \le V^{max}_j \quad \forall j, t
    \end{equation}

    \item \textbf{CDU Demand Satisfaction}: The total crude oil volume delivered to CDU $l$ must meet the production demand $DM_l$ for the planning horizon.
    \begin{equation} \label{eq:con_cdu}
    \sum_{j \in J} \sum_{t \in T} f^{BC}_{j,l,t} \ge DM_l \quad \forall l \in L
    \end{equation}
\end{itemize}

\paragraph{(3) Operational Constraints}
These constraints handle the logical coupling between continuous and binary variables (Big-M constraints) as well as physical limits.

\begin{itemize}
    \item \textbf{Unloading Connectivity}: Flow $f^{VS}$ is permitted only if vessel $v$ is in an active state ($x^W_{v,t}=1$) and a physical pipeline exists ($A_{vi}=1$).
    \begin{equation} \label{eq:con_unload_conn}
    f^{VS}_{v,i,t} \le F^{max}_{vs} \cdot x^W_{v,t} \cdot A_{vi} \quad \forall v, i, t
    \end{equation}

    \item \textbf{Charging Connectivity and Minimum Flow}: If charging tank $j$ feeds CDU $l$, the pipeline connection must be active ($z_{j,l,t}=1$), and the flow is typically constrained by a minimum pumping rate (semi-continuous flow).
    \begin{equation} \label{eq:con_cdu_conn}
    F^{min}_{bc} \cdot z_{j,l,t} \le f^{BC}_{j,l,t} \le F^{max}_{bc} \cdot z_{j,l,t} \cdot A_{jl} \quad \forall j, l, t
    \end{equation}

    \item \textbf{Transfer Flow Bounds}: Flow from primary to secondary tanks is limited by physical connections and pumping capacities.
    \begin{equation} \label{eq:con_transfer_conn}
    0 \le f^{SB}_{i,j,t} \le F^{max}_{bs} \cdot A_{ij} \quad \forall i, j, t
    \end{equation}
\end{itemize}
\end{suppinfo}

\bibliography{bib/ref.bib}

\end{document}